\documentclass[12pt,preprint]{aastex}

\def\dfplot#1{\plotone{#1}}
\def\BE{\begin{equation}}
\def\BEL#1{\begin{equation}\label{#1}}
\def\EE{\end{equation}}
\newcommand{\COBE}{{\it COBE}}

\newcommand{\LPH}{LPH~201.663+1.643}

\newcommand{\HII}{H\,{\scriptsize II}}
\newcommand{\OVI}{O\,{\scriptsize VI}}

\newcommand{\Halpha}{H$\alpha$}
\newcommand{\etal}{{\it et al.}~}

\newcommand{\degree}{^\circ}

\newcommand{\cm}{{\rm ~cm}}

\newcommand{\GHz}{{\rm ~GHz}}

\newcommand{\K}{{\rm ~K}}

\newcommand{\microK}{\mu{\rm K}}

\begin{document}

\title{Microwave ISM Emission Observed by WMAP}

\author{Douglas P. Finkbeiner\footnote{Hubble Fellow, Henry Norris
Russell Fellow}}
\affil{Princeton University, Department of Astrophysics,
Peyton Hall, Princeton, NJ 08544}


\begin{abstract}
We investigate the nature of the diffuse Galactic emission in
the Wilkinson Microwave Anisotropy Probe (WMAP) temperature anisotropy
data.  Substantial dust-correlated emission is observed at all WMAP
frequencies, far exceeding the expected thermal dust emission in the
lowest frequency channels (23, 33, 41 GHz).  The WMAP team (Bennett
\etal) interpret this emission as dust-correlated synchrotron
radiation, attributing the correlation to the natural association of
relativistic electrons produced by SNae with massive star formation in
dusty clouds, and deriving an upper limit of 5\% on the
contribution of Draine \& Lazarian spinning dust at K-band (23 GHz).
We pursue an alternative interpretation that much, perhaps most, of
the dust-correlated emission at these frequencies is indeed spinning
dust, and explore the spectral dependence on environment by
considering a few specific objects as well as the full sky average.
Models similar to Draine \& Lazarian spinning dust provide a good
fit to the full-sky data.  
The full-sky fit also requires a significant
component with free-free spectrum uncorrelated with \Halpha, possibly
hot ($\sim 10^6$K) gas within $30\degree$ of the Galactic center.

\emph{Subject headings: }
cosmic microwave background --- 
diffuse radiation ---
dust, extinction --- 
ISM: clouds --- 
radiation mechanisms: thermal --- 
radio continuum: ISM 
\end{abstract}

\section{INTRODUCTION}

\label{sec_intro}

The first year of data from the Wilkinson Microwave Anisotropy Probe
(WMAP) has sparked a revolution in the study of cosmology and the
early universe (Bennett \etal\ 2003, Hinshaw \etal\ 2003, Spergel \etal\ 2003).
In addition, it has provided full sky maps of the interstellar medium
at $23-94\GHz$ at a sensitivity of $\sim 200\microK$ per $7'$ pixel at
resolutions much higher than previous full-sky observations. 
The quality of the data has rekindled the debate about the origin of 
``anomalous'' dust-correlated microwave emission
(Kogut \etal\ 1996, de Oliveira-Costa 1997,1998,1999,2002,2003; Finkbeiner
\etal\ 2002, 2003; Banday \etal\ 2003; Lagache 2003) dubbed
``Foreground X'' by de Oliveira-Costa \etal\ (2002). 

The WMAP team relied on a Maximum Entropy Method (MEM) analysis to
determine that spinning dust\footnote{The idea of grain
rotational emission was first
discussed by Erickson (1957). More recently, after the discovery of
the population of ultra small grains, Ferrara \& Dettmar (1994) noted
that the rotational emission from such grains may be observable, but
their treatment incorrectly assumed Brownian thermal rotation of
grains.}
with the Draine \& Lazarian (1998) cold neutral
medium (CNM) spectrum accounts for less than 5\% of the
emission in any WMAP waveband.  We
show that for different models of spinning dust, these
constraints do not apply.  Indeed, in most of the Galaxy, spinning
dust emission (or some other mechanism such as magnetic
dipole emission, Draine \& Lazarian 1999) explain the data just as
well. 

We emphasize that the WMAP team's primary objective was to prevent
Galactic foreground emission from contaminating their cosmological
results based on the CMB anisotropy.  They have done a superb job of
modeling the foregrounds, regardless of whether the physical
interpretation of the subtracted temperature anisotropy signal is
correct.  
The WMAP polarization data also provide important cosmological
constraints as well as additional information about foregrounds. 
For example, if the Bennett \etal\ hard synchrotron component can be
shown to be $\sim30$\% polarized, then spinning dust would become an
unlikely explanation (see Lazarian \& Draine 2000).  On the other
hand, if it is unpolarized, the synchrotron would be ruled out.
Something between the two extremes is likely, and this will help us
determine the relative contributions of synchrotron and spinning
dust.  As the polarization data are not yet public, discussion of
polarization will be deferred to a later paper.

\section{ANALYSIS}

\label{sec_analysis}

A disadvantage of the MEM analysis used by the WMAP team is that it
can only find the components
it is searching for.  Significant spectral variation is expected not
only in the synchrotron spectrum, but also in any spinning dust that
might be present.  The WMAP strategy, using data at the WMAP
frequencies and priors based on the Haslam \etal\ (1982) 408 MHz map, the
Finkbeiner (2003) \Halpha\ map, and the Finkbeiner \etal\ (1999;
hereafter FDS99) dust
prediction, produces a ``synchrotron'' template highly correlated with
thermal dust emission (see Fig. 4 in Bennett \etal\ 2003).  The
explanation for this is that the most energetic electrons are found
nearest dusty star-forming regions, so there is a natural correlation
between dust and synchrotron emission.  This is certainly plausible,
but it is not the only possible explanation for the strong correlation
with dust. 

Instead, we start by asserting that there are three components whose spectral
characteristics are well enough understand to use external data sets,
subtract these off in fits with very little freedom, 
and then approach the remaining emission with an open mind.  These
three components are free-free, soft synchrotron, and thermal dust.  Here we
differentiate between ``soft'' synchrotron (traced by the Haslam map
at 408 MHz) and ``hard'' synchrotron as described by Bennett \etal\
(2003).  The existence of the soft synchrotron is not in dispute.  

\subsection{Free-Free}
Although warm ionized gas is
well traced by \Halpha\ emission in three recent surveys 
(VTSS, Dennison \etal\ 1998;
SHASSA, Gaustad \etal\ 2001;
WHAM, Reynolds \etal\ 2003) 
as presented by
Finkbeiner (2003) and corrected for dust using the Schlegel,
Finkbeiner \& Davis (SFD; 1998) dust map, one must worry about several
systematics.  Bennett \etal\ (2003) point out several sources of
uncertainty, paraphrased as follows:
(a) \Halpha\ light scatters off dust, while microwave free-free
scattering is negligible; (b) \Halpha\ calibration is
uncertain at the 10\% level; (c) for the 1/4 of the sky with no WHAM
coverage, there is significant uncertainty $(\sim1 R)$ separating
geocoronal \Halpha\ from Galactic ISM emission; (d) The assumed Case B
Balmer atomic rates may not be universally valid; (e) The ratio of
\Halpha\ to free-free microwave emission depends on $T_e$, which may
vary from object to object (e.g. Heiles \etal\ 2000); (f)
There is uncertainty in the ionization state of He; (g) ionized gas
with bulk velocity outside of the WHAM velocity coverage is not
included in the Finkbeiner map; and (h) the extinction correction
assumes that the ionized gas and dust are co-extensive along the line
of sight.  Of these sources of error, (e) and (h) likely dominate,
while (b), (c), and (f) are well characterized, and (a) and (d) are
probably small.

In spite of these difficulties, we can still make use of an \Halpha\
template.  The first step is to limit the \Halpha\ extinction
predicted by the SFD (1998) dust map to be $A < 2$ mag.  For dust with 
optical depth $\tau = A/(1.086$ mag) uniformly mixed with \Halpha\ emitting
gas
the observed intensity is 
\BE
I_{\nu,obs} = \frac{I_\nu}{\tau}\int_0^\tau d\tau'e^{-\tau'}
=\frac{I_\nu}{\tau}(1-e^{-\tau})
\EE
In the limit of large $\tau$, the intensity is simply reduced by a factor
of $\tau$.  For $A = 2$ mag the uniform mixing case results in a
reduction factor of 2.2, between the extremes of dust in front
(factor of 6.8) and dust behind gas (factor of unity).  Though uniform
mixing appears to be a reasonable approximation in general, we shall
bear in mind that factor of $\sim2$ errors in the free-free correction are
likely in some parts of the sky. 

A systematic variation in $T_e$ is also expected, with lower
temperatures toward the Galactic center, where higher metallicity
allows for higher cooling efficiency. 
Lacking an electron temperature map, we assume $T_e=5500\K$
everywhere, and use the
free-free ``haze'' (\S \ref{sec_haze}) to absorb the errors in this. 
Using the conversion formula in Valls-Gabaud (1998), an EM of 1
cm$^{-6}$ pc corresponds to 0.391 $T_4^{-1.017} 10^{-0.029/T_4}$ Rayleigh,
($T_4 = T_e/10^4$ K) which equals 0.636 R for $T=5500\K$.

Once an emission measure (EM) is derived for a given
pixel on the sky, the spectrum is constrained by physics (Spitzer
1978) so only one free parameter ($T_e$) is needed for the whole sky.
This is valid between the plasma frequency cutoff and the
optically thick limit at low frequencies.  According to catalogs
(Kurtz, Churchwell, \& Wood 1994) of optically thick ultra-compact \HII\
regions the filling factor of such regions is so small that at WMAP
frequencies a negligible fraction of the free-free is expected to be
optically thick, and certainly within our rather conservative mask
there can be no significant ultra-compact \HII\ regions. 


\subsection{Soft Synchrotron}
Synchrotron emission from relativistic electrons with energy
distribution $N(E) \sim E^{-\gamma}$ produces a synchrotron flux
density spectral index $\alpha = -(\gamma-1)/2$.  In temperature units
the index is $\beta = -(\gamma+3)/2$. 

The electron energy distribution is expected to vary spatially in the
Galaxy, with a harder spectrum near the sources of energetic
electrons, e.g. supernova remnants, and a softer spectrum elsewhere.
  Bennett \etal\ (2003)
argue that the natural correlation of SNR with dusty star-forming
clouds should cause a strong correlation between hard synchrotron
emission (at $\sim 30$GHz) and dust, with the correlation lessening at
lower frequencies, so that at 408 MHz, the Haslam (1982)
``soft'' synchrotron map shows very little correlation to dust, other
than the usual concentration of emission in the Galactic plane.  This
is certainly a cogent argument that synchrotron emission should become
more dust-correlated at higher frequencies, but provides no
evidence that such radiation dominates other mechanisms, such as
spinning dust.  

There is little doubt, however, that soft synchrotron (traced by the
408 MHz survey) is present, and important at least at K band (23 GHz)
so we include it in the removal of ``known components''.  A brightness
temperature index of $\beta=-3.05$ removes the most obvious K-band
synchrotron features at high latitude, so we assume this index is
constant throughout the high-latitude sky.

\subsection{Thermal Dust}
Another component whose morphology and (approximate) spectral
characteristics are understood is thermal (vibrational) dust emission
from grains large enough to be in equilibrium with the interstellar
radiation field.  Emission from this dust peaks at $\sim 140\micron$
and deviates strongly from a thermal blackbody spectrum.  A
Rayleigh-Jeans emissivity function of $\nu^2$ has often been assumed
in the literature (e.g. Draine \& Lee 1984; Schlegel, Finkbeiner \&
Davis 1998) but when dust temperature variation is accounted for, the
COBE FIRAS data (Fixsen \etal 1998) are better fit by a steeper power
law emissivity ($\beta=2.6$) near the peak and $\beta=1.7$ at lower
frequencies, with a break at about $500\GHz$ (Finkbeiner \etal 1999; FDS99).
This fit ties the IRAS and DIRBE data to FIRAS via a fit with only 4
global parameters describing the two emissivity laws, and the
requirement that the emission is dominated by grains in equilibrium
with the interstellar radiation field.  Predictions at $6'$ resolution
based on this fit are available on the web.\footnote{http://skymaps.info}

Because the spectral slope used to extrapolate the FDS99 thermal dust
spectrum through the WMAP bands is relevant to this work, we briefly
review its justification. 
The FDS99 two-component model yields a substantially better fit (reduced
$\chi^2=1.85$ compared to 31 for a $\nu^2$ model) even when the
spatial and spectral covariance of the FIRAS data are included (FIRAS
Explan. Supp., Brodd
\etal\ 1997).  The FIRAS data cannot differentiate between a single
component with non-power-law emissivity and two independent power law
components, but the physical reality of the two separate components is
not implausible.  Amorphous
silicates with a wide range of emissivity indices $\beta\sim 1.2-2.7$
have been observed in the lab (Agladze \etal\ 1996), including
amorphous MgO$\cdot$2SiO$_2$ which has a very high microwave
emissivity to optical absorption ratio, leading to rather different
mean temperatures (9K and 16K) for the two components.  These lab
emissivities were measured at $\sim300$ GHz and 20K and may become
steeper at lower frequencies.  However, this interpretation of the
spectral break is hardly unique; if the dominant emitter has such a
break in its emissivity function at 500 GHz, then a single component
could explain the data just as well.  Another explanation that has
been advanced is very cold dust grains spatially mixed with the warm
dust -- (Reach \etal 1995), though a physical mechanism
for keeping the grains so cold is not provided.  Such a
model would predict a steeper slope at lower frequencies
also. 

Regardless of interpretation, the Finkbeiner \etal\ (1999) model
has been very successful in the sub-mm - microwave, though small but
interesting deviations from the model have been observed by BOOMERANG
(150, 240, 410 GHz; Masi \etal\ 2001).  At lower frequencies, however,
there is a surprise.  

Comparing FDS99 predictions to COBE DMR, Finkbeiner \etal\ found that
COBE 90 GHz was slightly higher, but at 53 and 31 GHz the emission per
dust column is a factor of 2.2 and 31 higher than expected.  These
results are similar to the earlier Kogut \etal\ (1996) results derived without
an explicit dust temperature correction.  Because of this it was
expected that the FIRAS-based predictions would agree well with WMAP
94 GHz, but be significantly contaminated by some other dust-correlated
emission mechanism at lower frequencies. This appears to be true. 

Even though we can expect the vibrational thermal dust
brightness temperature to go as $T\sim\nu^\beta$ with $\beta=2$
through the WMAP channels, there may be other emissivity enhancements
such as magnetic dipole emission.  These are treated as a separate
component, so that temperature and other environmental dependence can
be searched for also.

The WMAP team (Bennett \etal\ 2003) 
has determined that the spectral slope from 94 to
61 GHz is $\beta=2.2$, significantly steeper than the
$\beta=1.7$ extrapolated from the Finkbeiner \etal\ 
(1999) fit.  However we find in this work that $\beta=1.7$ is
consistent with the data, and lacking any good theoretical reason to
depart from the FDS parameterization, we continue to use this index in
the following.

\subsection{Spinning dust template}
For a spinning dust template we use an empirically determined power of
dust color temperature (derived from the DIRBE $100/240\micron$ ratio)
times column density.  This template reflects the observed $T_{dust}$
dependence of the anomalous emission, but assumes constant spectral
dependence across the sky.  We use $T^2_{dust}$ times the FDS
prediction at 94 GHz, which is proportional to dust column times
$T^3_{dust}$ because the Rayleigh-Jeans dust prediction already
contains one power of temperature.  This modulation by $T^3_{dust}$
varies the dust template by a factor of $\sim 2$ peak-to-peak for the
SFD temperature range ($16-20$K).  There
is no obvious theoretical reason for this temperature dependence
other than the vague notion that hotter dust tends to be in a
more exciting radiation field, or perhaps has a size distribution with
more small grains, either one of which might enhance spinning dust
emission.  We will fit this average spectrum to the (nearly) full sky
and then examine a few regions of interest to see deviations from it.

\subsection{CMB anisotropy}
The CMB dipole is subtracted by the WMAP team, and used to calibrate
the anisotropy pattern to $<1\%$ uncertainty.  The WMAP sensitivity is
high enough that CMB anisotropy is clearly visible in all bands and
(for this work!) represents a significant nuisance background signal.
Because the cosmological signal is not correlated with the foreground
morphology for any physical reason, the cross correlation between the
foreground templates and the CMB (due only to chance alignment of
features) is minimal.  Nevertheless, we use the superposition of the 5
bands suggested by Bennett \etal (2003) to estimate the CMB
anisotropy, and subtract it from each band.  The superposition used is
\BEL{equ_cmb}
\Delta T_{CMB} = .109K -.684Ka -.096Q + 1.921V -.250W
\EE
where ($K,Ka,Q,V,W$) refer to the thermodynamic $\Delta T$ in
the 5 WMAP bands respectively. 
This works quite well
except where bright Galactic point sources near the ecliptic plane show the
beam asymmetries.\footnote{ It is often said that the inherently
asymmetric WMAP beams are symmetrized by scanning each point on the
sky from many different approach angles.  This is true near the
ecliptic poles, but at low ecliptic latitude the beam asymmetry is
obvious in the K and Ka band maps.}
This superposition only imperfectly cancels the foregrounds, of
course, allowing them to contaminate the CMB estimate and leak back
into our foreground analysis.  We find that any such contamination is
negligible, as it would appear as a foreground with the spectrum
of the CMB, and no such residual is found in our analysis. 

\subsection{Mask}
\label{sec_mask}
We mask out the 208 sources listed in the WMAP source catalog, and
also the LMC, SMC, M31, Orion-Barnard's Loop, and NGC 5090 (a total of
2.7\% of the sky).
Additionally, areas where the \Halpha\ extinction given by SFD is
greater than 2 mag, i.e. $A$(\Halpha) $\equiv 2.65 E(B-V) > 2$, are
masked (8.9\%).  The mask is smoothed to $1\degree$ FWHM to match the smoothed
data, excluding a total of 11.4\% of the sky from the analysis

\begin{figure}[tb]
\epsscale{1.0}
\dfplot{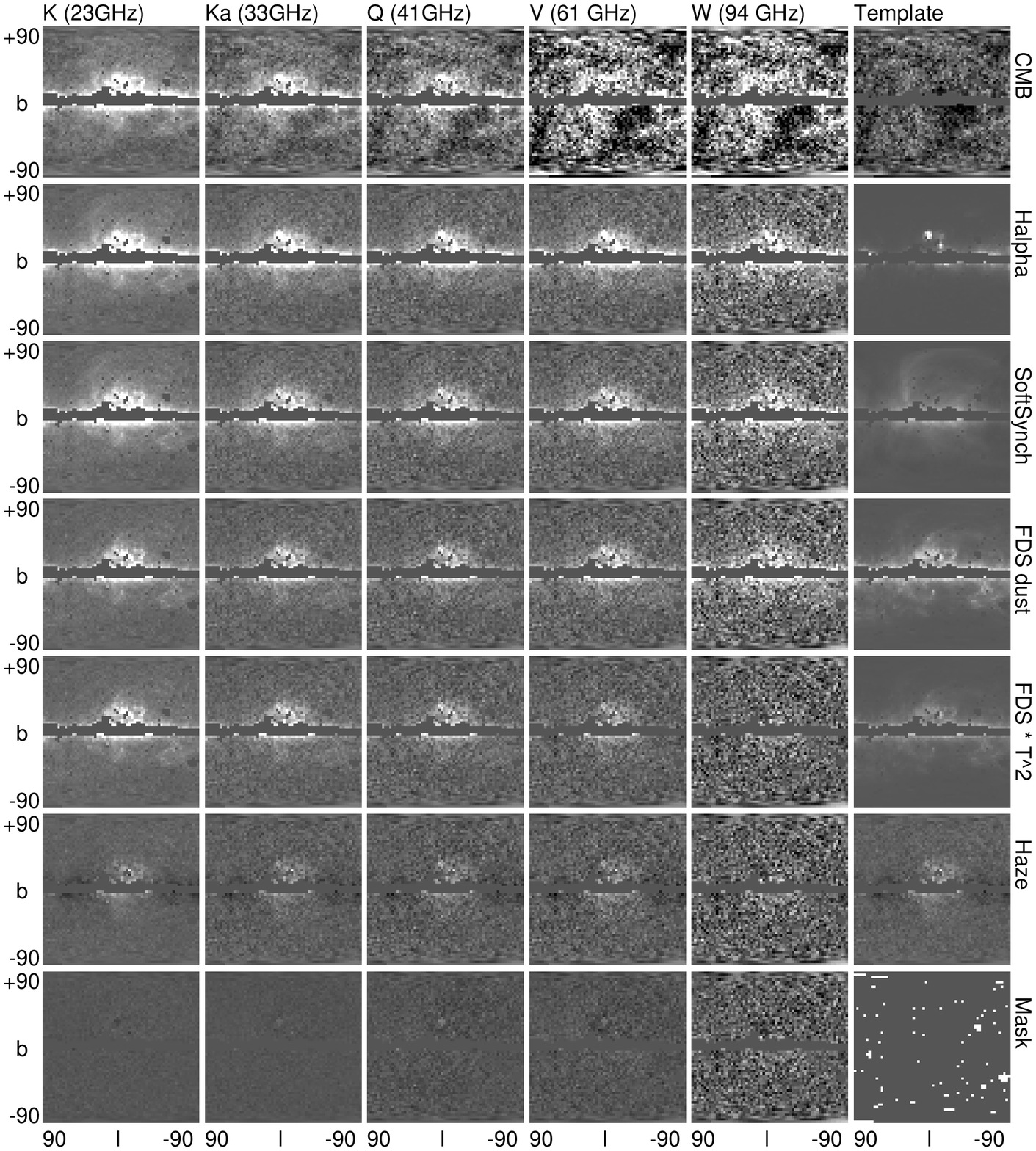}
\figcaption{The WMAP foreground grid; see detailed discussion in
\S\ref{sec_thefit}.
\label{fig_grid}
}
\end{figure}

\subsection{The Fit}
\label{sec_thefit}
The fit minimizes $\chi^2$ in an 8-dimensional parameter space over
unmasked pixels.  First all data and templates
are smoothed to $1\degree$ resolution using spherical harmonic
convolution on the $N=512$ HEALPix\footnote{see
http://www.eso.org/science/healpix.
Smoothing is done with spherical harmonic convolution on the HEALPix
sphere, using smoothing code implemented in IDL by D. Finkbeiner.}
sphere.  
Noise is derived from $\sigma_0$ values and $N_{obs}$, the number of
observations in each pixel (Jarosik \etal\ 2003). 
Initial guesses are determined by 
extensive experimentation with the data, then passed to the
simple-minded but fairly robust amoeba routine from Numerical
Recipes (Press \etal 1992) as implemented in IDL.\footnote{
IDL is a product of Research Systems, Inc. http://rsinc.com}
Although the fit is a simultaneous regression on all parameters, 
it is instructive to ``peel away'' the foregrounds one at a time and
show each WMAP band at each step. 

Figure \ref{fig_grid} is a 6 column by 7 row grid of images, each a Cartesian
projection of half the sky $(-90 < l < 90\degree; -90 < b < 90\degree)$. 
The first five columns are the WMAP bands in frequency order (K, Ka, Q,
V, W) = (23, 33, 41, 61, 94 GHz).  The sixth column is the template
being subtracted from the current row to yield the images on the row
below.  The first column is stretched so that -1 mK is black and 1 mK
is white ($=\pm 15$ Jy/sr).  Columns 2-4 are stretched to the same
intensity limits (in Jy/sr) as col. 1, and col. 5 matches the stretch
of column 4 in thermodynamic temperature units (to avoid saturation of
the gray scale).  Therefore, the CMB appears to become stronger from
columns 1 through 4, but is the same brightness in col. 5 as col. 4.  

Row 1: the sky signal (after CMB dipole removal) for the 5 WMAP
bands, with the plane and point sources masked.  Col. 6 is the CMB map
described above. 

Row 2: CMB-subtracted sky; note the \HII\ region around $\zeta$Oph,
above the plane and just left of center.  This feature is prominent in
the \Halpha\ based free-free template, shown at right.  The free-free
spectrum is constrained by physics, so only one parameter is fit for
an overall normalization. 

Row 3: Sky minus CMB and \Halpha; note the ``Loop I'' synchrotron
feature in the north, visible in the Haslam-based soft synchrotron
template at right.  A single power law is fit from the Haslam survey
frequency (408 MHz) to WMAP.  The spectrum is likely concave downward
over such a large range in frequency, but WMAP 23 GHz dominates the
fit, and is the only frequency for which this component is important
anyway. 

Row 4: Sky minus CMB, \Halpha, and soft synchrotron.  Dust correlated
emission is evident, as traced by the Finkbeiner \etal\ (FDS; 1999)
template at right.  This template is subtracted with amplitudes
determined by the FDS99 fit to FIRAS, not by cross-correlation with
the WMAP data.  Note that the subtraction at 94 GHz appears to be
nearly perfect. 

Row 5: Sky minus CMB, \Halpha, soft synchrotron, and FDS dust.  There
is still significant dust-correlated emission present at lower
frequencies.  The FDS dust emission template times $T_{dust}^2$ (SFD dust
temperature squared, based on DIRBE 100 and $240\micron$ emission ratio)
appears to be a good template for the remaining component, tentatively
interpreted as spinning dust, and is
shown at right.  A cross-correlation coefficient is determined independently
for each WMAP band. 

Row 6: Residual, after subtracting CMB, \Halpha, soft
synchrotron, FDS dust, and the putative spinning dust, is nicknamed
``the Haze.'' It looks nearly the same in every band and has the
spectrum of free-free emission.  A template for the Haze is produced
from the average fit residuals in K and Ka bands, and subtracted from
all five WMAP bands with a free-free spectrum.  There are no free
parameters in this operation, other than one value per sky pixel
determined from the K and Ka maps. 

Row 7: The WMAP residuals after subtracting all above templates show
very little remaining other than measurement noise.  The one exception
is residual emission from $\zeta$Oph which appears to peak at 41 GHz
and roll off to either side (relative to free-free).  The point source
mask used in the analysis is shown in the lower right hand panel.

\section{INTERPRETATION}
The three important results of this analysis are that (1) the quality
of the fit is very high, (2) the dust-correlated component has a
reasonable spectrum to be spinning dust, and (3) a ``haze''
unassociated with \Halpha\ appears to be present in the data.

\begin{figure}[tb]
\epsscale{0.75}
\dfplot{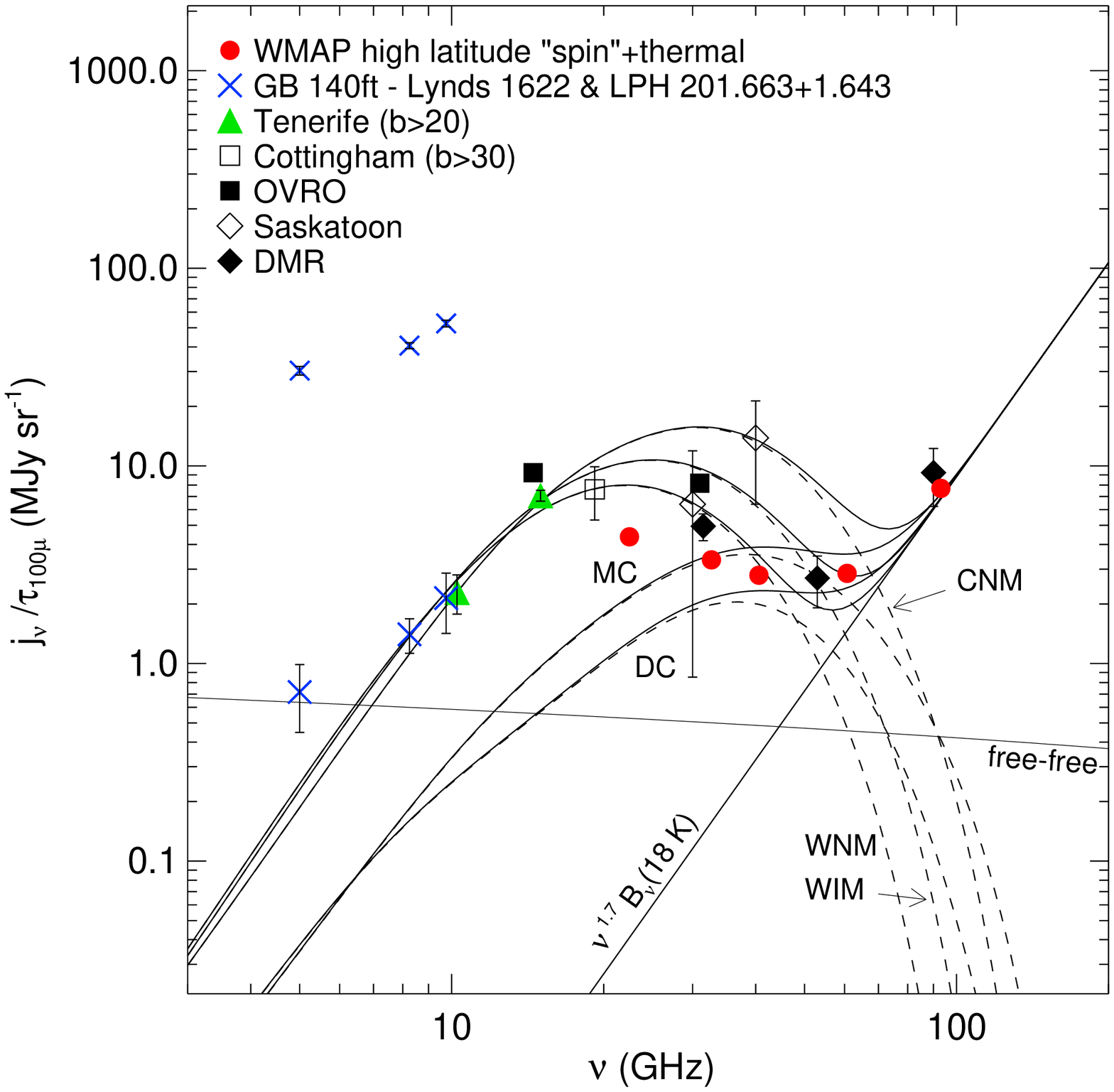}
\figcaption{Model dust emissivity per $\tau_{100\mu}$ for DC, MC, CNM, WNM, and
WIM conditions (as in Draine \& Lazarian 1998b, Figure 9)
with (\emph{solid lines}) and without (\emph{dashed lines})
contribution from vibrational dust at mean temperature. 
For comparison with the Draine \& Lazarian models, unit
$\tau_{100\mu}$ corresponds to $N$(H) = $2.13\times10^{24}\cm^{-2}$. 
Gray line is emission from free-free for 
$<n_en_p>/<n_H>=0.01\cm^{-3}$ averaged along the line of sight. 
WMAP data from Table 2 (including thermal dust) are overplotted
(\emph{filled circles})
Also shown are measurements from the \COBE/DMR (\emph{solid diamonds})
from Finkbeiner \etal\ (1999), similar to Kogut
\etal\ (1996); Saskatoon (\emph{open diamonds}) (de Oliveira-Costa 
\etal\ 1997); the Cottingham \& Boughn $19.2\GHz$ survey 
(\emph{open square}) (de
Oliveira-Costa \etal\ 1998), OVRO data (\emph{solid squares}) 
(Leitch \etal\ 1997);
Tenerife data (\emph{triangles})(de Oliveira-Costa \etal\ 1999);
GB 140 foot (\emph{crosses})(Finkbeiner \etal 2002).
The OVRO points have been lowered a factor of 3 relative to Draine \&
Lazarian (1998b, Figure 9), because the unusual dust temperature near
the NCP caused an underestimate of $\tau_{100\mu}$ along those lines
of sight.  
Given the large range of model curves, all measurements 
are consistent with some superposition of spinning dust, vibrational
dust, and free-free emission. 
\label{fig_dl_spinfig}
}
\end{figure}

\begin{figure}[tb]
\epsscale{0.75}
\dfplot{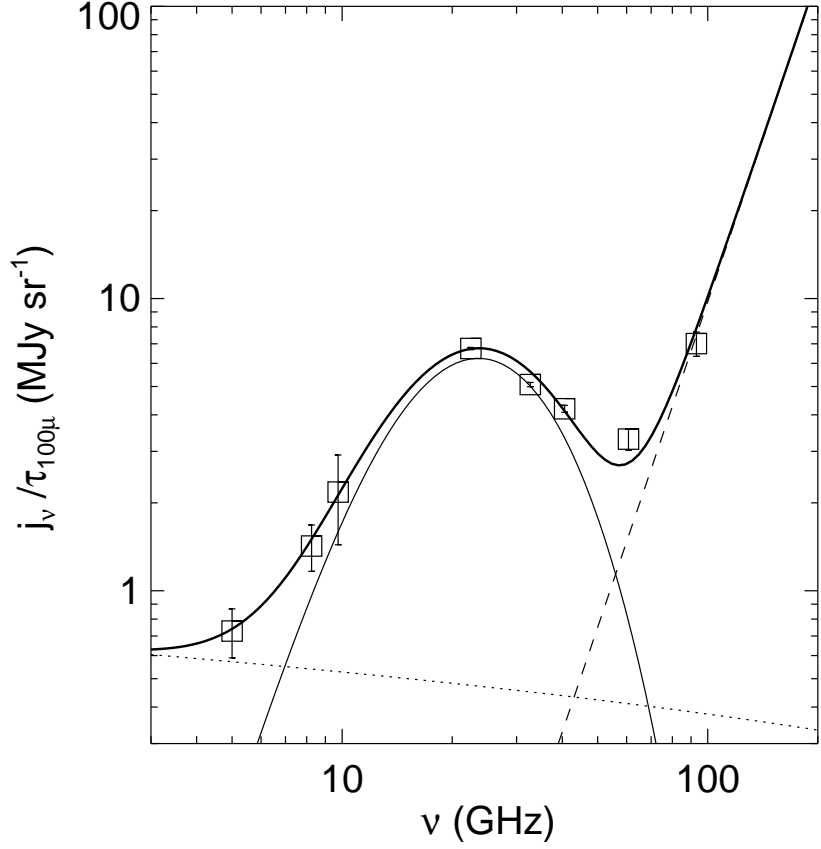}
\figcaption{
LDN1622.  Green Bank (5, 8, 10 GHz) measurements in a $6'$ beam (Finkbeiner et
al. 2002) and WMAP (23-94 GHz) data points smoothed to a $60'$ beam
(\emph{symbols}), each with statistical $1\sigma$ error bars only.
Thermal dust estimate from Finkbeiner \etal\ (1999)
(\emph{dashed line}), free-free estimate (\emph{dotted line}),
spinning dust (\emph{thin line}), and sum of free-free, thermal
dust, and spinning dust (\emph{thick line}).  The spinning dust model
is a superposition of two Draine \& Lazarian models: $0.6\times$ WNM
plus $0.11\times$ CNM model. 
\label{fig_dl_spinfig_l1622}
}
\end{figure}

\begin{figure}[tb]
\epsscale{0.75}
\dfplot{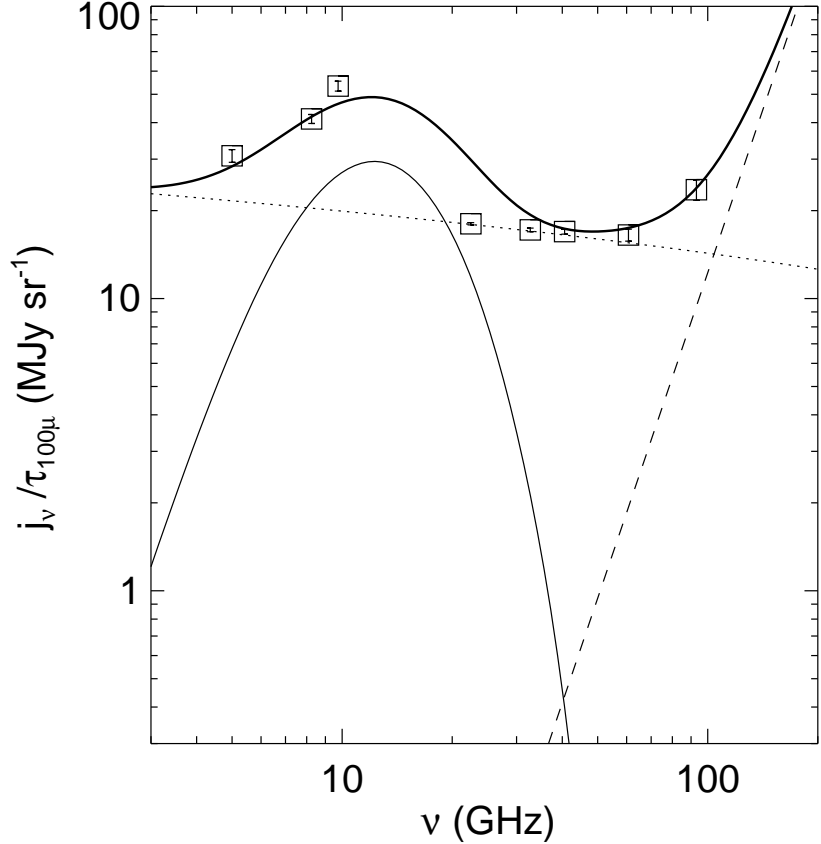}
\figcaption{
\LPH.  Green Bank (5, 8, 10 GHz) and WMAP (23-94 GHz) data points
(\emph{symbols}), thermal dust estimate from Finkbeiner \etal\ (1999)
(\emph{dashed line}), free-free estimate (\emph{dotted line}),
spinning dust (\emph{thin line}), and sum of free-free, thermal
dust, and spinning dust (\emph{thick line}).  
None of the published spinning dust models come close to fitting this object. 
For reference, the CNM model is shown, shifted in frequency and
amplitude.  This curve may not represent a physical spinning model
for any reasonable choice of parameters. 
Green Bank points are $6'$ and WMAP is smoothed to $60'$, so deviation
of the fit from WMAP points may be due to beam mis-match
between Green Bank and WMAP, but
the outlier at 10 GHz is significant. 
\label{fig_dl_spinfig_lph}
}
\end{figure}

\begin{figure}[tb]
\epsscale{1.0}
\dfplot{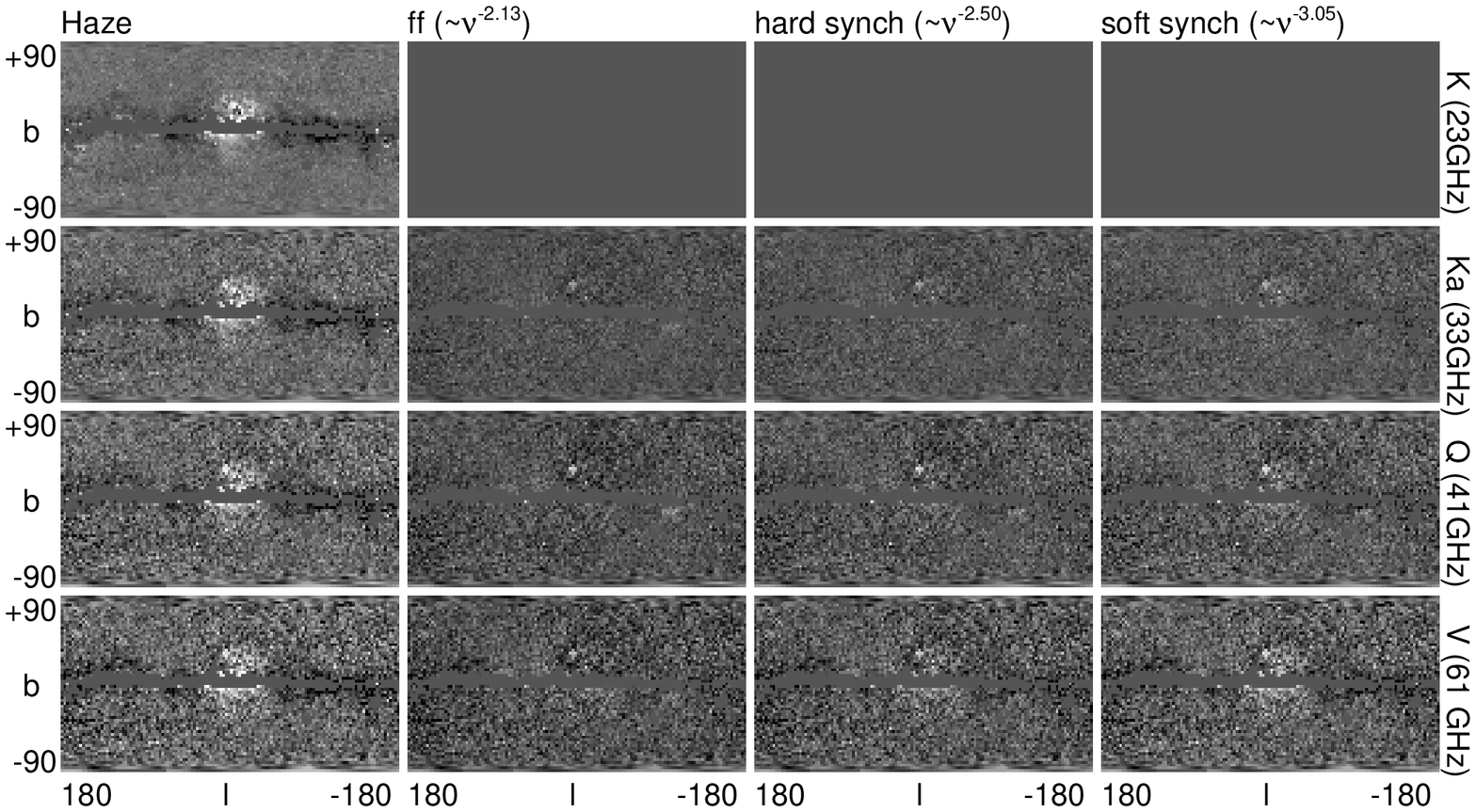}
\figcaption{The Haze is determined in 4 WMAP bands by subtracting 
CMB, soft synchrotron (Haslam template), free-free (\Halpha\ template)
and spinning dust.  Using the K-band haze as a template, it is then
subtracted from Ka, Q, and V bands assuming various power laws. 
A free-free spectrum fits most of the sky well, apart from the $\zeta$
Oph cloud $(l,b) = (5\degree,25\degree)$.
\S\ref{sec_haze}.
\label{fig_haze}
}
\end{figure}

\subsection{Quality of fit}
The quality of the fit is remarkable.  All
that is subtracted from each sky map is a CMB template; a map with
free-free spectrum (the \Halpha\ based template plus the haze);
soft synchrotron, which is hardly controversial at low frequencies and
unimportant at high frequencies; FDS dust emission extrapolated from
the FIRAS fit; and emission strictly
correlated with FDS dust times $T_{dust}^2$.  This last component is
the only one that can contain hard synchrotron emission or spinning
dust emission -- the others are all well constrained and physically
understood.  The surprisingly small residuals imply that the physical
mechanism primarily responsible for this emission is closely tied to
the dust grains themselves (e.g. spinning dust) rather than a
mechanism relying on the spatial correlation of dust and some other
emission source (e.g. hard synchrotron emission). 

In doing this fit, we have not allowed the spectral dependence of any
component to float from pixel to pixel, as the WMAP team MEM analysis
did (Bennett \etal\ 2003).  
Though it may seem overly conservative to fix the spectral dependence
of each component, this rigidity in the fit allowed the discovery of
the free-free haze, and the quality of the resulting fit indicates
that spectral variation of the components is not required to fit WMAP
data off the plane. 
Because the
templates were applied directly rather than as priors to a MEM
analysis, evaluation of $\chi^2$ is straightforward.  Values 
given in Table 1 are for $1\degree$ FWHM gaussian beams; this reduces
measurement noise to approximately $15\microK$ per beam in each band. 
Because of the CMB and haze subtraction, the residual in each band
is actually a superposition of the foreground templates and 5 WMAP
bands appropriately weighted.  The smoothed WMAP noise for each band
is then combined in a weighted quadrature sum to produce the expected
noise in the 5 residual maps.  This process assumes that the 5 WMAP
bands have uncorrelated noise, and that the \Halpha, Haslam, and FDS
templates have no measurement noise. 
The fact that $\chi^2$ per dof is of
order unity at high latitude (Table 1) indicates that these
assumptions are justified in Q, V, and W bands, and less so in K and
Ka bands.  At $|b|<30\degree$ the $\chi^2$/dof is formally poor (13.6
at K and 7.8 at Ka, and 2 at Q) yet represents removal of 98\%, 97\%,
and 99\% of the
variance in the WMAP data respectively.  It is significant that these
values are with respect to the $1\degree$ FWHM smoothed data; the
residual variance of $60\microK$ per beam at K band is less than half
of the median measurement noise per pixel, and relatively smaller in
the other bands.  

The residuals in the Bennett \etal\ (2003) analysis are also
impressively small. 
It is impossible to compare the results of this study
to the Bennett \etal\ MEM analysis, which fit several numbers for each
pixel on the sky (constrained by prior templates).  Such a MEM
analysis does not allow for an easy interpretation of $\chi^2$, so
a smaller fit residual would not imply the superiority of one
foreground parameterization over the other.  In fact, the same freedom
that allows the MEM to minimize foreground contamination of the
cosmological signal may actually \emph{inhibit} the physical
interpretation of the foreground signal by allowing an overly simple
model to fit the data.

\subsection{Spinning Dust Spectrum}
Draine \& Lazarian (1998, 1999) chose to phrase the comparison between
theory and observation in terms of emissivity per H atom, using units
of Jy cm$^2$ sr$^{-1}$ per H in their work\footnote{
Plots in the early Draine \& Lazarian articles are incorrectly labeled 
``Jy sr$^{-1}$ per H'', an error repeated by Finkbeiner \etal\
(2002).
}.  This ratio is appealing to theorists
because it directly relates the dust model to chemical abundances in
the ISM.  However, measuring the hydrogen column density, including molecular,
atomic, and ionized forms, is technically difficult. 
Finkbeiner \etal\ (2002) measured cross-correlation slopes of
emission against the SFD98 dust map, and converted to $N($H$)$ using 1
mag $E(B-V)$ = $8\times 10^{21}$ $N($H$)$.  Although this conversion is
appropriate for diffuse high latitude material, the gas to dust ratio
is known to vary in general, and phrasing measurements in terms of
$E(B-V)$ or $N($H$)$ when they have nothing to do with either has
led to confusion.  

In this work, we stay as close as possible to the empirical data and
state emissivity per dust in terms of optical depth at $100\micron$,
or $\tau_{100\mu}$.  This optical depth is defined with respect to a
blackbody at the reference dust temperature $T_{ref}=18.175$ K used in
SFD98.  This temperature is the median high-latitude dust color
temperature determined from the DIRBE $100\micron$ / $240\micron$
intensity ratio assuming a $\nu^2$ emissivity law.  For the SFD
calibrations, $\tau_{100\mu}$ of unity corresponds to 266 mag $E(B-V)$
or $2.13\times 10^{24}$ $N($H$)$, so the reader may convert Draine \&
Lazarian models and previously published measurements to $j_\nu /
\tau_{100\mu}$ in Jy sr$^{-1}$ by multiplying by $2.13\times 10^{24}$. 

The cross-correlation coefficients for the FDS99 $\times T^2_{dust}$
spinning dust template are given in Table 2.  These may be added to
the FDS99 thermal dust prediction, where the sum is taken at dust
temperature $T_{ref}=18.175$ K.
In Figure \ref{fig_dl_spinfig} the dust-correlated emission per
$\tau_{100\mu}$ is shown for 5
Draine and Lazarian (1998) models of electric dipole emission from
spinning dust, along with the cross-correlation points from WMAP and
other projects.  Note that the thermal dust emission and ``spinning
dust'' emission have been combined for WMAP, just as they have been
for other experiments (intentionally or not).  The WMAP data points
fall about a factor of two lower than the WNM model.  These WMAP
points represent the average high-latitude sky, so there is no reason
to expect them to agree with any particular Draine \& Lazarian model,
but rather a superposition.

Two spinning dust candidates previously studied with the Green Bank
140 foot telescope (Finkbeiner \etal 2002) are also shown, Lynds Dark
Nebula 1622 (Figure 3, see also Hildebrand \& Kirby 2003) and
\LPH\ from the Lockman, Pisano, \& Howard (1996) catalog of diffuse
\HII\ regions (Figure 4).  A superposition of the WNM and CNM models
plus a small amount of free-free yields a reasonable fit
for LDN1622. 
Note that the thermal dust level is not being fit, merely extrapolated
from the far IR according to the FDS99 FIRAS fit.  Spinning dust
models have a few free parameters to tune, and some models may come
closer to fitting the LPH cloud, though this will require further
theoretical work.  Lacking a physical model for this object, a curve
(the Draine \& Lazarian WNM model shifted down a factor of 1.8 in
frequency and up a factor of 3.7 in amplitude) is shown in Figure 4
for comparison.  Neither of these objects would be well fit by a
superposition of 
synchrotron, free-free, and thermal dust without including spinning
dust.  

\subsection{The Haze}
\label{sec_haze}
A rather substantial amount of emission (both positive and negative)
was attributed above to the ``free-free haze,'' without any interpretation.
As the spectrum of this component most resembles free-free (Figure
\ref{fig_haze}), the
straightforward interpretation is that it constitutes the
error in the \Halpha\ based free-free template.  Because the spatial
template is orthogonal to the others by construction, it contains
positive and negative parts.  
Some error is to be
expected both from variation in warm gas temperature about the mean
$T_{gas} = 5500$K used, and from the fact that some gas is hot enough
$(> 10^6$K) that the \Halpha\
recombination line is strongly suppressed.  This need for hot gas
applies especially to the region south of the Galactic center, where there
is significant emission seen in the haze but none of the other
templates have a similarly extended feature of significant
brightness. 

 Further study of this diffuse haze, perhaps using \OVI\
absorption measurements toward distant stars below the Galactic
Center, may be able to verify this hypothesis about its origin. 
We emphasize that the tentative identification of the haze as
free-free is based only on its spectral shape, and 
if WMAP polarization data show a significant polarization, that would
suggest very hard synchrotron ($\sim\nu^{-0.1}$) emission as a source.
Further consideration of the haze is deferred to
another paper. 

A HEALPix map of the K-band haze template used in Figure 1, 
defined as $(T_K+2.186\ T_{Ka})/2$, is available on the web\footnote{
http://skymaps.info}. 

\section{CONCLUSIONS}
We have performed an analysis of the WMAP temperature anisotropy data,
using free-free, soft synchrotron, FDS thermal dust, and spinning dust as
templates.  We find:

1.  The FDS thermal (non-magnetic) dust prediction is good on average
    over the high-latitude sky at 94 GHz, and
    even in dark cloud LDN1622 and \HII\ region \LPH. 

2.  Free-free is fit by $T = 5500$K ionized gas (rather cool, but see
    e.g. Heiles \etal\ 2000), plus a haze
    toward to the Galactic center with free-free spectrum.   This is
    the first detection of emission from this component.  Further
    study is needed to test whether this component is indeed
    free-free or perhaps a very hard synchrotron component. 

3.  The residual dust-correlated emission is well fit by $T^3_{dust}
    \times$ dust column density over most of the sky.  Such a model,
    along with free-free, soft synchrotron, and FDS thermal dust, accounts
    for $\sim98\%$ of the variance in the WMAP data. 

4.  There is no conflict between the general idea of spinning dust
    emission and the WMAP data.  In spite of the Bennett \etal\ (2003)
    statement that the Draine \& Lazarian CNM model accounts for less
    than 5\% of the ISM emission (which is true), spinning dust
    remains a reasonable hypothesis both morphologically and
    spectrally.

5.  The Green Bank Galactic Plane Survey GPA data at 8 and 14 GHz
    (Langston \etal\ 2000) rule out hard synchrotron as a
    majority component in the diffuse ISM in the WMAP bands (see
    companion paper,
    Finkbeiner, Langston, \& Minter 2003).  This leaves spinning dust
    as a viable alternative. 

The WMAP data alone cannot establish the relative levels of spinning
dust, magnetic dust, and hard synchrotron emission,
since any of these theories can be adjusted to explain the data.
However, objects such as LDN1622 (Figure 3) are clearly not dominated
by synchrotron in the WMAP bands, and are well fit by a superposition
of spinning dust models. 
Therefore, the dominance of spinning dust at 23 GHz is a good
working hypothesis.  Further data (including surveys at lower
frequencies, better WMAP signal to noise,
and WMAP polarization data coming soon) will help lift the remaining
degeneracies and determine the identity of the dust-correlated
microwave emission from``Foreground(s) X.''


I am indebted to Bruce Draine, Ed Jenkins, and David Schlegel for
encouragement and helpful advice.  
Roger Hildebrand called the WMAP spectrum of LDN1622 to my attention,
and provided other insights. 
Lyman Page, Chuck Bennett, and Gary Hinshaw provided helpful comments
on the proper use of the WMAP data in this work.  Comments from an
anonymous referee substantially improved the text. 
Some of the results in this paper were derived using HEALPix
(G\'orski, Hivon, and Wandelt 1999). 
This research made use of the NASA Astrophysics Data
System (ADS) and the IDL Astronomy User's Library at
Goddard\footnote{http://idlastro.gsfc.nasa.gov/}.  DPF is a Hubble
Fellow supported by HST-HF-00129.01-A.
\newpage


\bibliographystyle{unsrt}
\bibliography{gsrp}


\clearpage
\begin{deluxetable}{l|rrrrrrr}
\footnotesize
\tablewidth{0pt}
\tablecaption{Chi squared results, single band
   \label{table_chisq1}
}
\tablehead{
\colhead{map} &
\colhead{$b$ cut} &
\colhead{K} &
\colhead{Ka} &
\colhead{Q} &
\colhead{V} &
\colhead{W} &
\colhead{Mean}
\\
\colhead{} &
\colhead{deg} &
\colhead{23 GHz} &
\colhead{33 GHz} &
\colhead{41 GHz} &
\colhead{61 GHz} &
\colhead{94 GHz} &
\colhead{}
\\
\colhead{} &
\colhead{(2)} &
\colhead{(3)} &
\colhead{(4)} &
\colhead{(5)} &
\colhead{(6)} &
\colhead{(7)} &
\colhead{(8)}
}
\startdata
W       &  all  &391.8645 &139.7677 & 93.6773 & 43.6718 & 31.5367 &140.1036 \\
W       & $<30$ &583.3032 &206.6596 &133.5638 & 54.6387 & 38.4774 &203.3285 \\
W-C     &  all  &122.9073 & 17.7629 &  8.7606 &  7.0739 &  2.2803 & 31.7570 \\
W-C     & $<30$ &185.5578 & 27.3719 & 13.7034 & 10.6546 &  2.8084 & 48.0192 \\
W-C-T   &  all  & 14.4250 &  2.8392 &  1.8906 &  1.7544 &  1.1566 &  4.4132 \\
W-C-T   & $<30$ & 27.1182 &  4.6733 &  2.7873 &  2.5053 &  1.2265 &  7.6621 \\
W-C-T-H &  all  &  9.4000 &  5.4792 &  1.6167 &  1.4913 &  1.1282 &  3.8231 \\
W-C-T-H & $<30$ & 13.6420 &  7.8132 &  2.0239 &  1.8904 &  1.1632 &  5.3065 \\
\hline
W-C-T-H & $>45$ &  2.0515 &  1.5306 &  1.0962 &  1.1211 &  1.0938 &  1.3787 \\
\enddata
\tablecomments{
Col. (1):
W=WMAP data,
C=CMB map defined in Eq. \ref{equ_cmb},
T=Template fits, 
H=Haze defined in \S\ref{sec_haze},
Col. (2): Galactic latitude cut, in addition to mask defined in
\S\ref{sec_mask},
Col. (3-7): $\chi^2$ per dof for the 5 WMAP bands in $1\degree$ beams,
Col. (8): mean $\chi^2$ for the 5 bands.
In W band the $\chi^2$ is dominated by imperfect
CMB removal.  In other bands imperfect foreground removal is
dominant, but the model reduces $|b|<30\degree$ excess variance (above
measurement
noise) in the data by factors of 46, 30, 130, and 60 in K-V bands
respectively. 
}
\end{deluxetable}

\begin{deluxetable}{l|rrr}
\footnotesize
\tablewidth{0pt}
\tablecaption{Correlation slopes
   \label{table_corrslp}
}
\tablehead{
\colhead{Band} &
\colhead{Freq.} &
\colhead{Spinning Dust} &
\colhead{Spin + Thermal}
\\
\colhead{} &
\colhead{GHz} &
\colhead{MJy sr$^{-1}\tau_{100\mu}^{-1}$} &
\colhead{MJy sr$^{-1}\tau_{100\mu}^{-1}$}
\\
\colhead{(1)} &
\colhead{(2)} &
\colhead{(3)} &
\colhead{(4)}
}
\startdata
K  &  22.5 &  4.33391 & 4.37490 \\
Ka &  32.7 &  3.18627 & 3.34376 \\
Q  &  40.6 &  2.45289 & 2.79613 \\
V  &  60.7 &  1.39604 & 2.85618 \\
W  &  93.1 &  0.88224 & 7.69205 \\
\enddata
\tablecomments{
Col. (1):
WMAP band name,
Col. (2):
effective central frequency for free-free spectrum,
Col. (3): cross-correlation slope of spinning dust component per
$100\micron$ optical depth,
Col. (4): sum of spinning dust and thermal (vibrational) dust
predicted by Finkbeiner \etal\ (1999) per $\tau_{100\mu}$.  These data
are plotted in Figure \ref{fig_dl_spinfig}. 
}
\end{deluxetable}
\clearpage

\end{document}